# DEVELOPMENT OF 3-CELL TRAVELING WAVE SRF CAVITY*

F. Furuta[†], V. Yakovlev, T. Khabiboulline, K. McGee, Fermi National Accelerator Laboratory, Batavia, USA
R. Kostin, P. Avrakhov, Euclid Techlabs, Bolingbrook, IL, USA

## Abstract

Traveling wave SRF cavity is a new technology and requires a multi-stage process for development. Conceptual designs have been proposed to adopt TW resonance in an SRF cavity The early stages of developments have been funded by several SBIR grants to Euclid Techlabs which were completed in collaboration with Fermilab. A 3-cell proof-of-principle TW cavity was fabricated as part of that and demonstrated the TW resonance excitation at room temperature. A TW resonance control tuner for the 3-cell was also fabricated and the preliminary tests were performed. Now, the 3-cell cavity is being processed and prepared for the first cryogenic testing.

## INTRODUCTION

Niobium SRF cavities have a theoretical peak magnetic field which limits the accelerating field to 50 - 60 MV/m using standard available designs. Presently all SRF cavities operate in a standing wave (SW) resonance field in which particles experience an accelerating force alternating from zero to peak. Changing to a travelling wave (TW) mode operation can improve the efficiency of acceleration per cell defined as the transit time factor $T$ ($T=E_{acc}/E_{ave.}$, $E_{acc}$; accelerating gradient, $E_{ave}$; average field gradient over the cell gap) since a TW resonance field propagate along with a TW structure and particles in such resonance field can experience a constant acceleration force. A TW structure proposed in the early study showed a $T$ of 0.9 and the study suggested it could effectively increase the acceleration per cell more than 20% compared with a SW cavity [1]. Thereby, TW mode operation allows higher accelerating fields with niobium SRF cavities beyond the maximum gradient of 50 - 60 MV/m in a SW mode for the same peak surface magnetic field condition. This approach explores the path to 70 MV/m and higher using standard Niobium materials and processing technology. Increased cavity gradients can dramatically reduce the cost of SRF accelerators.

## 3-CELL TRAVELING WAVE CAVITY

### Early Achievements

Traveling wave (TW) SRF cavity is a new technology and requires a multi-stage process for development. Conceptual designs have been proposed to adopt TW resonance in an SRF cavity [1, 2]. The early stages of developments have been funded by several SBIR grants to Euclid Techlabs which were completed in collaboration with Fermilab through a 1-cell prototype and 3-cell TW cavity fabrication and testing. It demonstrated the feasibility of the fabrication and processing of multi-cell TW structure [3] and achieved the TW resonance excitation in the 3-cell TW cavity at room temperature [4, 5].

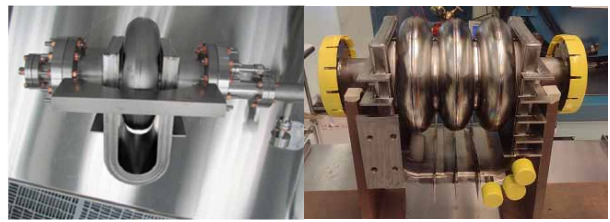

Figure 1: 1-cell prototype (left) [3] and 3-cell TW cavity (right).

### TW Resonance Control in the 3-cell Cavity

To achieve TW resonance at 2K in the 3-cell cavity, the following technical concepts were investigated and demonstrated during the early R&Ds; 1) the stiffened cavity design, 2) RF control using two input power couplers, and 3) a special tuner device [4, 5].

The loaded quality factor $Q_{load}$ at 2 K in VTS will be around $10^8$, making the cavity bandwidth very narrow and sensitive to microphonics. The stiffening ribs were welded on the waveguide to reinforce the resonator (Fig. 1 right). Based on the simulations, the stiffened cavity design and fine tuning of RF input signals would be enough to withstand microphonics detuning.

The waveguide of the 3-cell cavity has two RF input couplers and three RF pick-up probes. During the study at room temperature, "clockwise" and "anti-clockwise" traveling waves inside the cavity and waveguide loop were mathematically extracted from the pick-up probe signals. The "anti-clockwise" traveling wave was suppressed by RF power redistribution and phase control between the two input signals. Thus, the desired "clockwise" traveling wave was established in the 3-cell cavity at room temperature. Figure 2 shows RF feed and measurement scheme at room temperature for the 3-Cell TW cavity by Euclid [4].

Lorentz force compensation was also considered to maintain the TW resonance at 2 K VTS conditions. A special tuner device (the matcher) for the 3-cell cavity was designed and fabricated to attach on the cavity. A matcher will deform the waveguide wall and decouple partial modes to compensate the Lorentz force [5]. The preliminary test on the matcher test assembly (Fig. 3) at room and liquid nitrogen temperatures indicated the feasibility of achieving TW resonance in 2 K. Figure 4 shows 3D model of the 3-cell protype and the matcher.

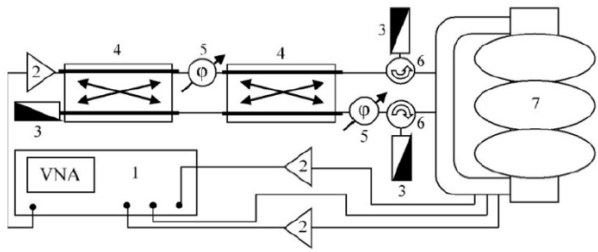

Figure 2: RF feed and measurement scheme at room temperature for the 3-Cell cavity: 1 - Vector Network Analyzer; 2 - power amplifier; 3 - matched load; 4 – 3 dB hybrid; 5 - phase shifter; 6 -circulator; 7 – the cavity [4].

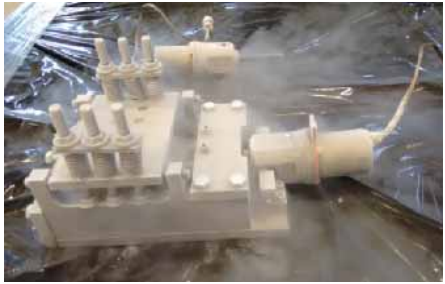

Figure 3: The matcher test assembly after $LN_2$ test.

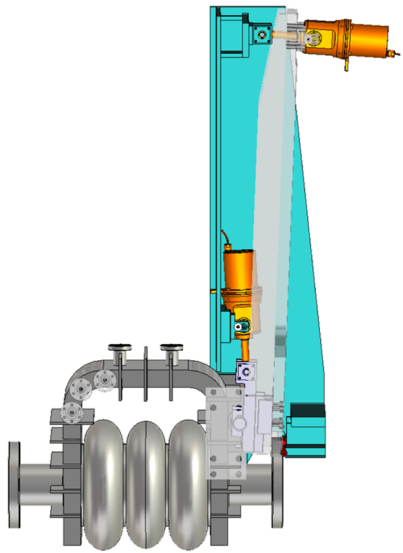

Figure 4: 3D model of the cavity with the matcher on WG.

## CRYOGENIC TESTING PREPARATIONS OF THE 3-CELL CAVITY

Activities on the 3-cell cavity were resumed last year toward the first cryogenic testing in collaboration with Euclid.

### Surface Processing

A 120μm Buffered Chemical Polishing (1:1:2, 48% HF: 69.5% $HNO_3$: 85% $H_3PO_4$) was applied first on the 3-cell cavity at Argonne National Lab (ANL) [6]. High Pressure DI water Rinsing (HPR) and a heat treatment in high-vacuum furnace (800 degC, 3hrs) were carried out at Fermilab. After additional 10μm BCP at ANL followed by HPR at Fermilab, the cavity was proceeded to cell tuning.

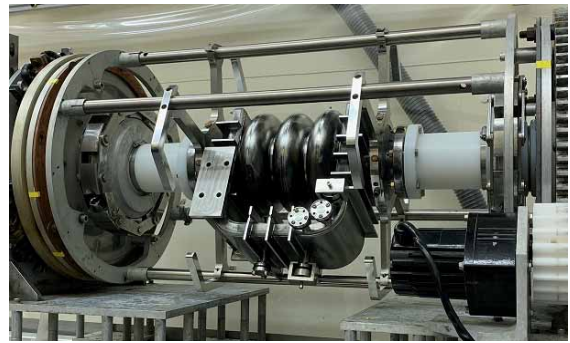

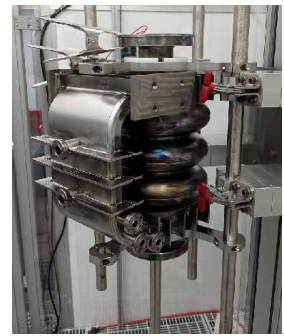

Figure 5: BCP arrangement at ANL (top), HPR arrangement at FNAL (bottom).

### Cell Tuning

Cell tuning hardware for the 3-cell cavity is designed and fabricated by Euclid Techlabs (Fig. 6). The 3-cell cavity was tuned in SW mode to the simulated field distributions from CST MWS [7] for 3 cell TW mode. The TW operation frequency should be close to the second eigenmode shown with a grey line in Fig. 7.

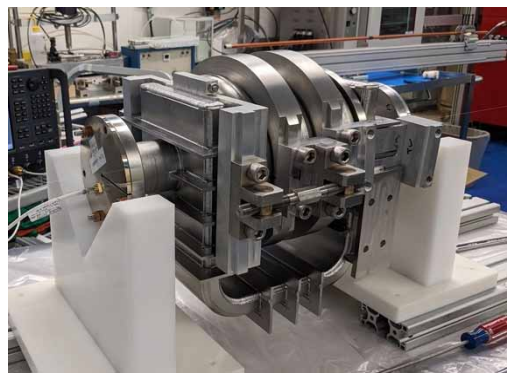

Figure 6: Tuning hardware on the 3-cell.

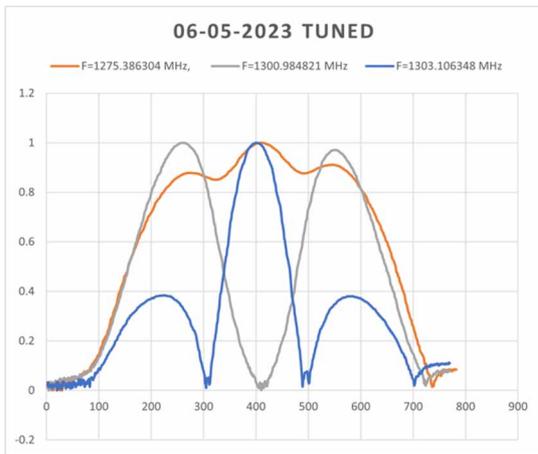

Figure 7: The field profiles post tuning in SW mode.

### TW Resonance Excitation at Room Temperature

Fig. 8 shows the arrangement in a clean room to excite TW resonance. An example of TW at 1301.06 MHz excited at room temperature after cell tuning was shown in Fig. 9 with two processed signals; blue shows a suppressed backward wave and magenta shows a maximum forward wave. Yellow is an unprocessed signal from the calibration pick up.

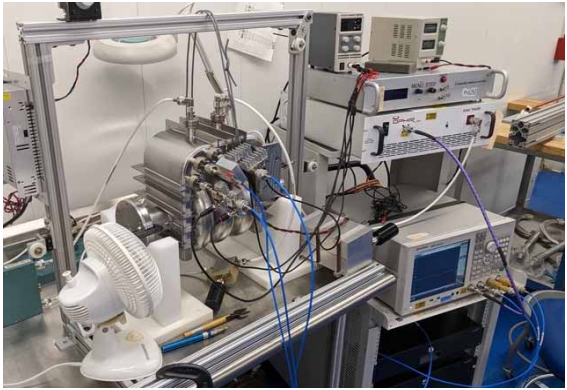

Figure 8: The cavity arrangement for TW resonance excitation at room temperature post cell tuning.

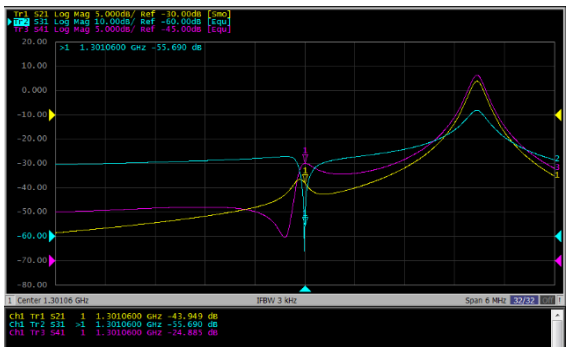

Figure 9: An example of TW signals.

### Status and Plans

Final machining and modification on some test hardware is in progress. Fig. 10 shows the model of VTS arrangement. The goals of TW 3-cell cavity cryogenic testing are to demonstrate TW resonance excitation in the 3-cell at 2K, study TW resonance control in cryogenic temperature, and study high gradient in TW regime.

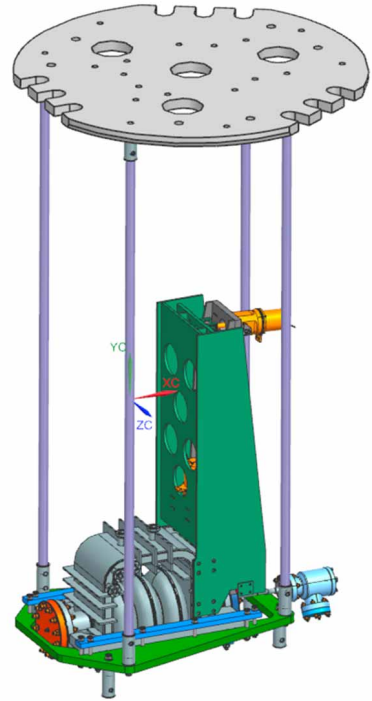

Figure 10: 3D model of the 3-cell prototype VTS arrangement.

## SUMMARY

Surface processing and cell tuning on the TW 3-cell cavity had been completed. TW resonance excitation at room temp was also confirmed. The cavity and the WG tuner are ready for cryogenic test assembling. Final machining and modification on the test hardware is in progress. The first cryogenic testing of the TW 3-cell cavity will be carried out soon.